\documentclass[12pt]{article}
\usepackage{graphicx}
\usepackage{amssymb}
\usepackage{amsmath}
\usepackage{amsfonts}
\usepackage{array,hhline,dcolumn} 
\usepackage{epstopdf}
\usepackage[dvips]{epsfig}

\begin{document}

\thispagestyle{empty}

\title{\vspace{-1.in} A Study of Detector Configurations \\
for the DUSEL CP Violation Searches \\
Combining LBNE and DAE$\delta$ALUS \\
}

\author{The DAE$\delta$ALUS Collaboration$^*$}

\maketitle

\hrule

\vspace{0.1in}

\noindent {\it Abstract:}
This study presents comparative CP sensitivities for various sets of
water Cerenkov and liquid argon detectors combined with various
running scenarios associated with DAE$\delta$ALUS and LBNE neutrino
beams at DUSEL.  LBNE-only running scenarios show fairly small differences in
sensitivity for the various detector combinations.  On the other hand,
the DAE$\delta$ALUS-only and DAE$\delta$ALUS-plus-LBNE running gives
significantly better sensitivity for a detector combination that
includes at least 200 kt of Gd-doped water Cerenkov detector,
exceeding the sensitivity of a Project-X 10 year run. A 300 kt Gd-doped water
Cerenkov detector yields the best sensitivity for combined running.

\vspace{0.1in}

\hrule

\vspace{0.5in}

This study examines the sensitivity of various design configurations
proposed for DUSEL for the physics of CP violation.  We rank-order ten
configurations of beams and detectors. 
We consider three types of detector ``units'' which are then arranged in
configurations consisting of sets of three units. The units are:

\begin{itemize}
\item \textbf{WCGd:}100 kt Gd-doped water Cerenkov detector with
 $\approx$20\% high quantum efficiency PMT coverage and/or light
 concentrators to realize good efficiency for the $\sim$5 MeV
 Cerenkov light signal expected from neutron capture on Gd \cite{ntag};

\item \textbf{WC:} ~~100 kt water Cerenkov detector modules with 15\% high
quantum efficiency PMT coverage;

\item \textbf{LAr:} ~17 kt of LAr.
\end{itemize}
We consider three types of neutrino sources, with 10 year running-periods:
\begin{itemize}
\item \textbf{LBNE alone} -- which is 30$\times10^{20}$ protons on target (POT) in neutrino mode
followed by 30$\times10^{20}$ POT in antineutrino mode.  This is the standard
10-year run-plan prior to the startup of Project X \cite{Gina}.

\item \textbf{DAE$\delta$ALUS alone} -- which is in antineutrino mode,
following the plan described in ref. \cite{EOI}.

\item \textbf{DAE$\delta$ALUS + LBNE} -- which is the standard plan
for DAE$\delta$ALUS antineutrino running combined with LBNE running for the full 10 years in \underline{only} neutrino mode.  These programs can proceed simultaneously.
\end{itemize}

The code used for this study is described in detail in
ref. \cite{EOI}. The DAE$\delta$ALUS event rates do not depend on the
mass hierarchy, but the LBNE rates do; for this study we use the
normal hierarchy.  For these comparisons, an input uncertainty of
$\delta(\sin^{2} 2\theta_{13})=0.005$ from the upcoming reactor
experiments, has been assumed \cite{EOI, firstpaper, huber}.

The assumptions concerning beam fluxes  
for DAE$\delta$ALUS and LBNE have been
described in ref. \cite{EOI}.  The DAE$\delta$ALUS flux assumes a
decay at rest (DAR) beam arising from the stopped pion decay chain
produced by an incident proton beam of 800 MeV.  The assumed locations
and powers of the sources are: 1 MW at 1.5 km, 2 MW at 8 km, and 5 MW
(when time averaged across the 2-phase run-plan) at 20 km.
The LBNE flux files used in this discussion are \cite{MaryPrivate}:
\begin{itemize}
\item dusel120e250i002dr280dz1300km\_flux.txt (neutrino flux) 
\item dusel120e250ni002dr280dz1300km\_flux.txt (antineutrino flux)
\end{itemize}
These files are for an 120-GeV proton-on-target, on-axis, NuMI-like
beam with a 280-m decay.  

Refs. \cite{EOI}, \cite{firstpaper} and \cite{Barger} describe the
assumptions related to signal events and backgrounds in the water
Cerenkov detector for both the DAE$\delta$ALUS and LBNE running.  We
assume a 67\% reconstruction efficiency for DAE$\delta$ALUS inverse
beta decay signal events in a WCGd.  We assume a 15\% reconstruction
efficiency for LBNE charged current quasielastic signal events in
either WC or WCGd.  DAE$\delta$ALUS event backgrounds arise from
beam-off sources, which can be measured during beam-off, and from
intrinsic $\bar \nu_e$ in the beam, which is measured by the near
accelerator.  The systematics on these backgrounds arise from the
statistical error on the measurements.  LBNE event backgrounds arise
from neutral current (NC) misidentification and intrinsic
electron-flavor neutrinos in the beam.  In the WC detector, we assume
a 10\% error on the LBNE NC and intrinsic backgrounds.

In this study, we also consider an LAr detector option.  Because this
target has no free protons, there is a low interaction rate for a
DAR beam, and this detector does not contribute to the DAE$\delta$ALUS event
sample.  However, this detector can efficiently observe interactions at LBNE
beam energies.  To address the fact that the LAr detector has better
signal to background for NC mis-identification, we scale the WC
mis-identification background for each LAr unit. The assumption is
that the overall efficiency for electron neutrinos or antineutrinos
for LAr is 90\%, which is 6 times higher than WC.  The efficiency for
accepting neutral current $\pi^0$ background is unchanged, so that
this effectively gives a reduction factor for neutral current $\pi^0$
background of 6 (to 17\% of the WC).  These scale factors are
consistent with the present assumptions of the LBNE collaboration
\cite{private} and other studies \cite{T2KLAr, T2KProp}.  To be specific, for this study, we assume:
\begin{itemize}
\item 3$\times$WC = 1.0*background,
\item 2$\times$WC + 1$\times$LAr = 0.72*background,
\item 1$\times$WC + 2$\times$LAr = 0.44*background,
\item 3$\times$LAr = 0.17*background,
\end{itemize}
where ``background''
refers to the mis-identification background for signal events produced
by the LBNE beam.

As a simple benchmark to compare scenarios, we choose a point in\\
$(\sin^{2} 2\theta_{13}, \delta_{CP})$ space and report the $1\sigma$
error on $\delta_{CP}$ for each configuration.  We have chosen
$(\sin^{2} 2\theta_{13}=0.05, \delta _{CP}=-90^{\circ})$. A comparison
of ten scenarios is made in Table 1, ranked according to increasing
sensitivity (1 is worst, 10 is best).

\bigskip
\begin{table}[t] \centering
\begin{tabular}
{|l|c| c|c|c|}\hline
Rank & Source  &   Configuration  &  1$\sigma$ error& Comment \\ \hline 
1.& DAE$\delta$ALUS alone &  1$\times$WCGd & 34$^\circ$ & Worst \\ 
2.&LBNE alone & 3$\times$WC &  25$^\circ$ & \\
3.&LBNE alone & 2$\times$WC+1$\times$LAr & 24$^\circ$ & \\
4.&LBNE alone & 1$\times$WC+2$\times$LAr & 23$^\circ$ & \\ 
5.& DAE$\delta$ALUS alone & 2$\times$WCGd & 22$^\circ$ & \\ 
6.&  DAE$\delta$ALUS + LBNE & 1$\times$WCGd+2$\times$WC & 18$^\circ$ &\\
7.& DAE$\delta$ALUS alone & 3$\times$WCGd & 17$^\circ$ & \\ 
8.&  DAE$\delta$ALUS + LBNE & 2$\times$WCGd+1$\times$WC & 15$^\circ$ & \\
9.&  DAE$\delta$ALUS + LBNE & 2$\times$WCGd+1$\times$LAr & 15$^\circ$ & \\
10.&  DAE$\delta$ALUS + LBNE & 3$\times$WCGd & 13$^\circ$ & Best\\ \hline
\end{tabular}
\caption{Comparison of Configurations for $\sin^2 2\theta_{13}=0.05$ and $\delta_{CP}=-90^\circ$}%
\end{table}%

In the case of running DAE$\delta$ALUS alone, only the number of units
of WCGd are relevant, because the WC and LAr
units are not sensitive to DAE$\delta$ALUS events.  Table 1 shows that
running DAE$\delta$ALUS alone with only 1 WCGd unit has poorest
sensitivity (rank=1).  As this configuration is statistics limited,
adding units of WCGd immediately increases DAE$\delta$ALUS'
sensitivity. Two units of WCGd are already better than running LBNE
alone (rank= 5) and 3 WCGd units is best for DAE$\delta$ALUS alone
(rank=7).  The 3 WCGd unit is also the best for the combined beams
(rank=10).

The sensitivities for running LBNE alone are very similar for the various
configurations (ranks = 2-4).  This is simply a statement that the LAr
sensitivity is designed to match the WC sensitivity by balancing target size
against efficiency. Three
configurations of WC and LAr units are provided.  
Note that WC and WCGd are equivalent in the case of LBNE.  This is 
because the LBNE beam energy is $>100$
MeV;  in this range neutron capture is no longer a relevant
tag.

The best sensitivities arise when the two beam sources are combined
with at least 200 kt of Gd-doped water Cerenkov detector.  For a
discussion of how the beams are complementary, resulting in this
substantial improvement in sensitivity, see refs. \cite{EOI} and
\cite{Agarwalla}.  For the combined source running, the case of 1 WCGd
module and 2 WC modules (rank=6) is significantly worse than the cases
of 2 WCGd modules and 1 WC or 1 LAr (ranks =8 and 9). As discussed
above, the best sensitivity arises from 300 kt of Gd-doped water
Cerenkov detector (rank=10).

The conclusions of Table 1 follow for all 
values of $ \delta_{CP}$.   To see this, we  
provide plots which show the comparisons across all values of $ \delta_{CP}$,
for $\sin^2 2\theta_{13}=0.05$, in Figures 1 to 4.

Figure 5 shows the fraction of non-zero (and non-180$^\circ$)
$\delta_{CP}$-space which can be sampled at 3$\sigma$.  For
simplicity, only the designs which ranked 6, 9 and 10 in Table 1 are
shown.  This allows comparison for the best cases for 100 kt, 200 kt
and 300 kt WCGd.   The Project X expectation is also shown.
This assumes a 300
kt set of WC detectors with the LBNE conventional beam for $10^{22}$
POT in neutrino mode and $10^{22}$ POT in antineutrino mode. Both the
200 kt and 300 kt designs, paired with the combined
DAE$\delta$ALUS-plus-LBNE-$\nu$-only neutrino source, exceed the
Project X expectation \cite{EOI, Agarwalla}.

In conclusion,  the reach of the combined LBNE and DAE$\delta$ALUS beams is
outstanding, when at least 200 kt of Gd-doped water Cerenkov detector
is included in the plan.  The 300 kt design provides the best sensitivity.

\begin{figure}[p]\begin{center}
{\includegraphics[width=5.5in]{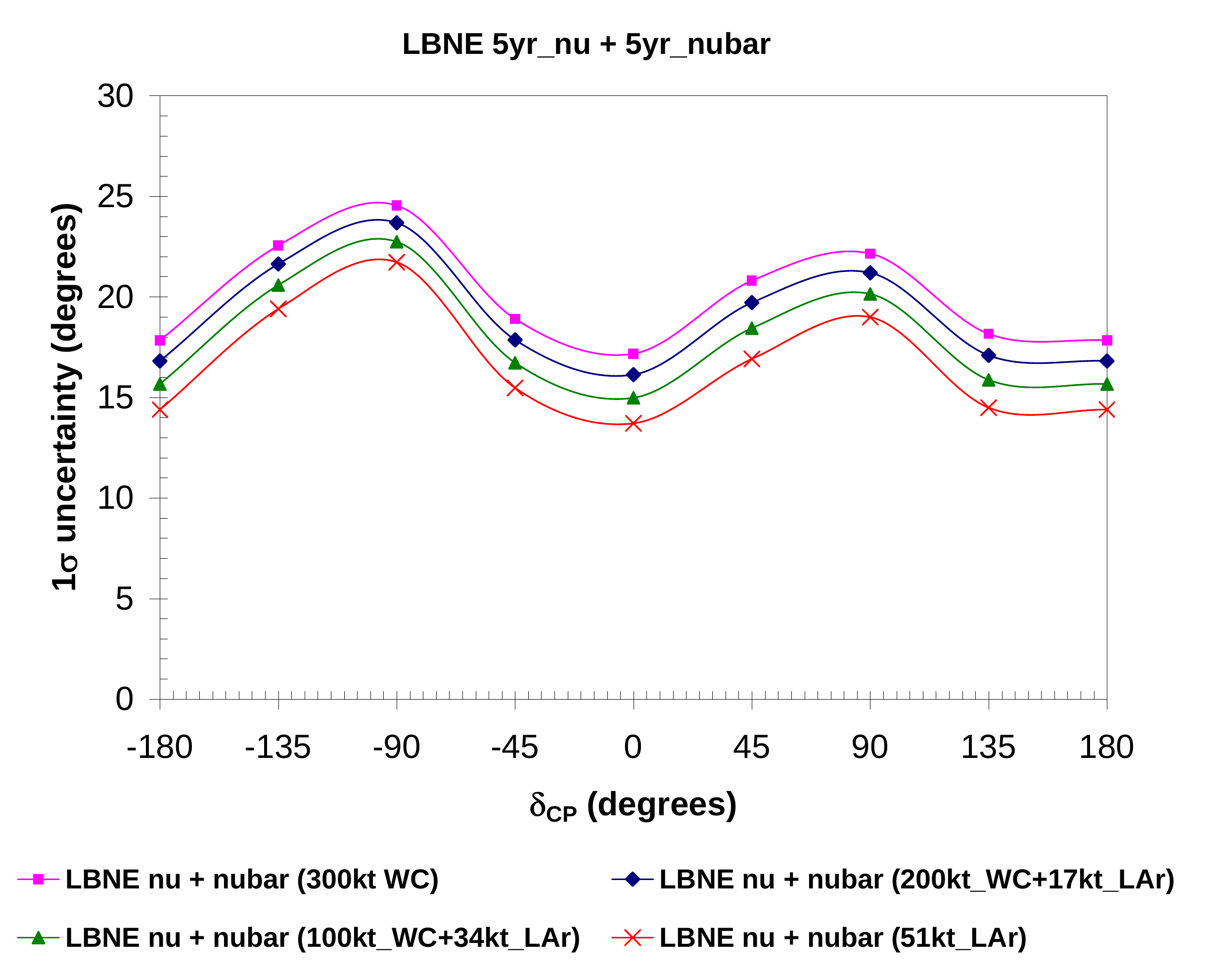}} 
\end{center}
\caption{The $\delta_{CP}$ measurement sensitivity at $1\sigma$ for LBNE-only running for various configurations
as a function of $\delta_{CP}$. Each measurement is for 5 years of $\nu$ plus 5 years $\bar{\nu}$ running.}
\end{figure}

\begin{figure}[p]\begin{center}
{\includegraphics[width=5.5in]{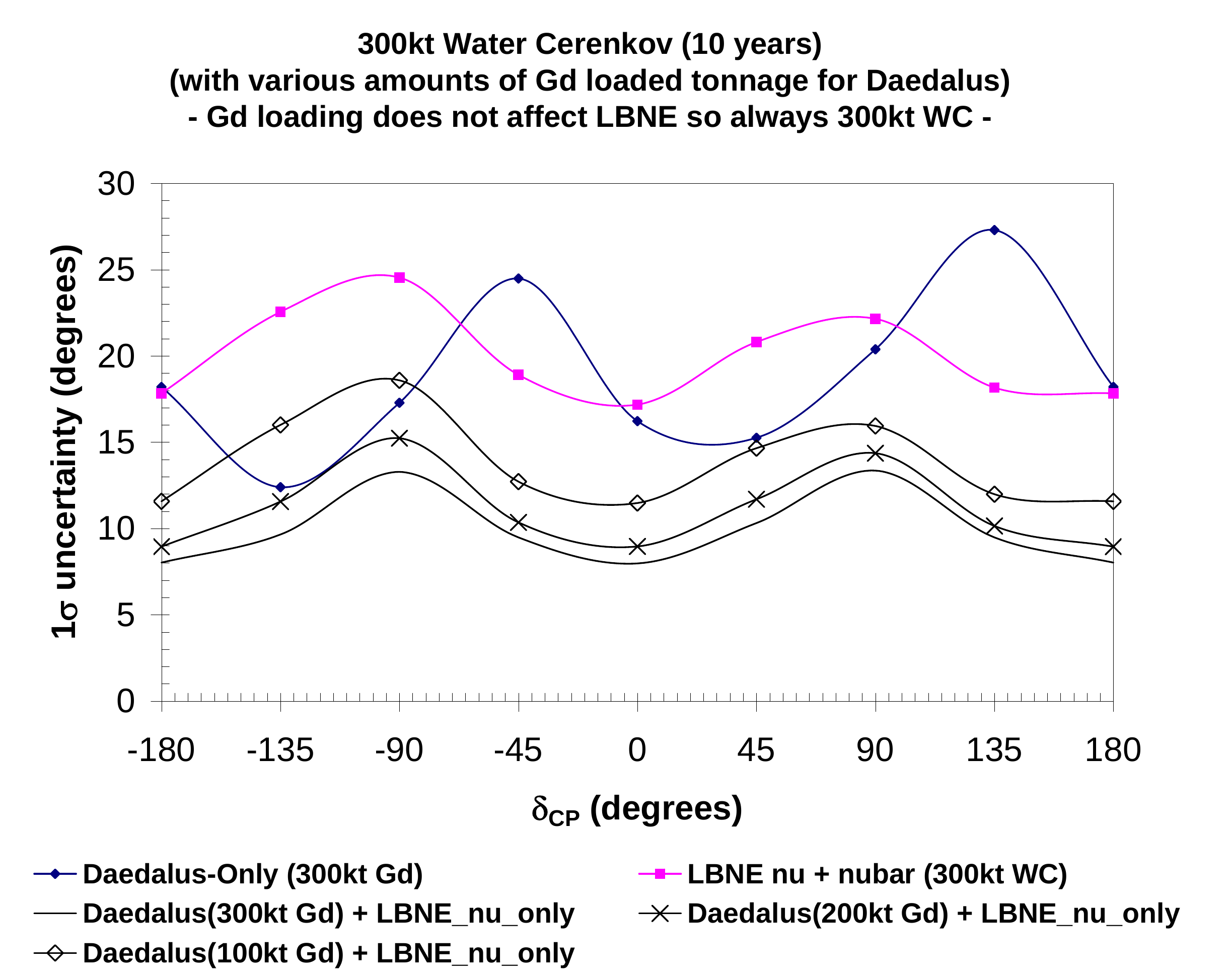}} 
\caption{The $\delta_{CP}$ measurement sensitivity at $1\sigma$ as a function of $\delta_{CP}$
for various configurations
involving units of WC and WCGd which total 300 kt.
The DAE$\delta$ALUS-only and DAE$\delta$ALUS plus LBNE $\nu-$only
scenarios are for 10 years of running and the LBNE-only is for 5 years of $\nu$ plus 5 years $\bar{\nu}$ running.} 
\end{center}

\end{figure}

\begin{figure}[p]\begin{center}
{\includegraphics[width=5.5in]{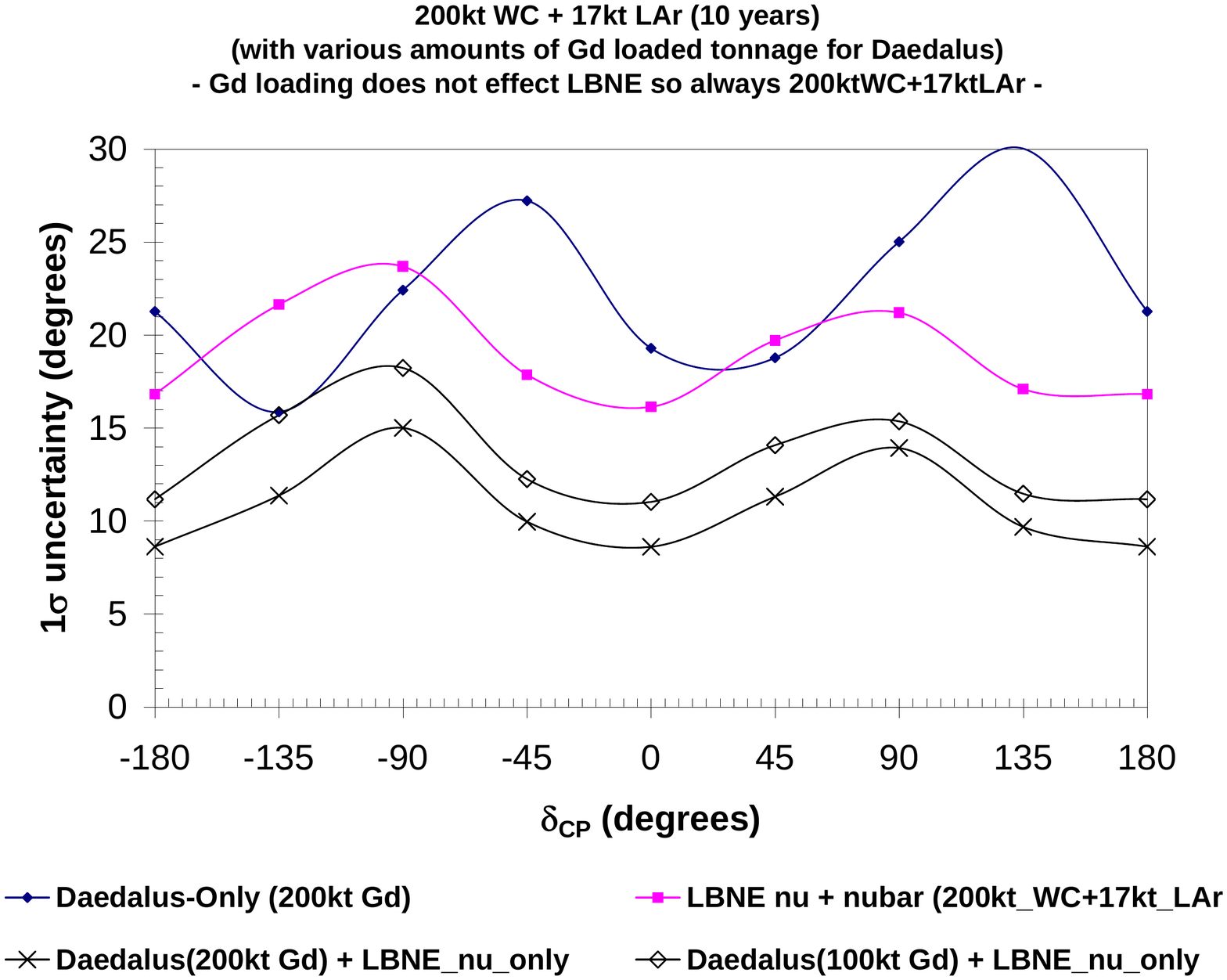}} 
\end{center}
\caption{The $\delta_{CP}$ measurement sensitivity at $1\sigma$
as a function of $\delta_{CP}$ for various configurations
involving units of LAr, WC and WCGd, with 200 kt of WCGd.
One example of only 100 kt of WCGd is also shown.
The DAE$\delta$ALUS-only and DAE$\delta$ALUS plus LBNE $\nu-$only
scenarios are for 10 years of running and the LBNE-only is for 5 years
of $\nu$ plus 5 years $\bar{\nu}$ running.} 
\end{figure}

\begin{figure}[p]\begin{center}
{\includegraphics[width=5.5in]{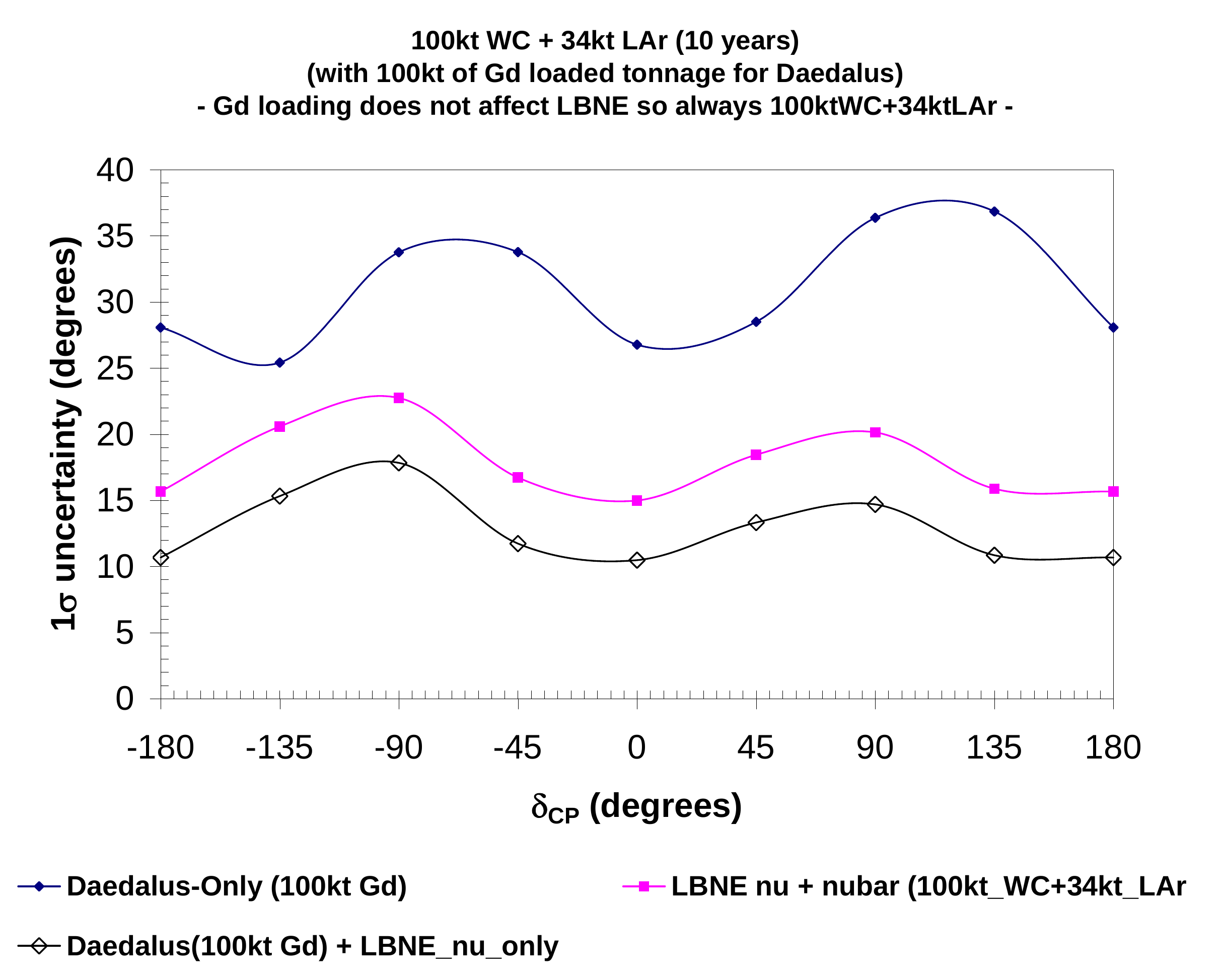}} 
\end{center}
\caption{The $\delta_{CP}$ measurement sensitivity at $1\sigma$
as a function of $\delta_{CP}$ for various configurations
involving units of LAr,  WC and WCGd, with only 100 kt of WCGd.
The DAE$\delta$ALUS-only and DAE$\delta$ALUS plus LBNE $\nu-$only
scenarios are for 10 years of running and the LBNE-only is for 5 years
of $\nu$ plus 5 years $\bar{\nu}$ running.} 
\end{figure}

\begin{figure}[p]\begin{center}
{\includegraphics[width=5.5in]{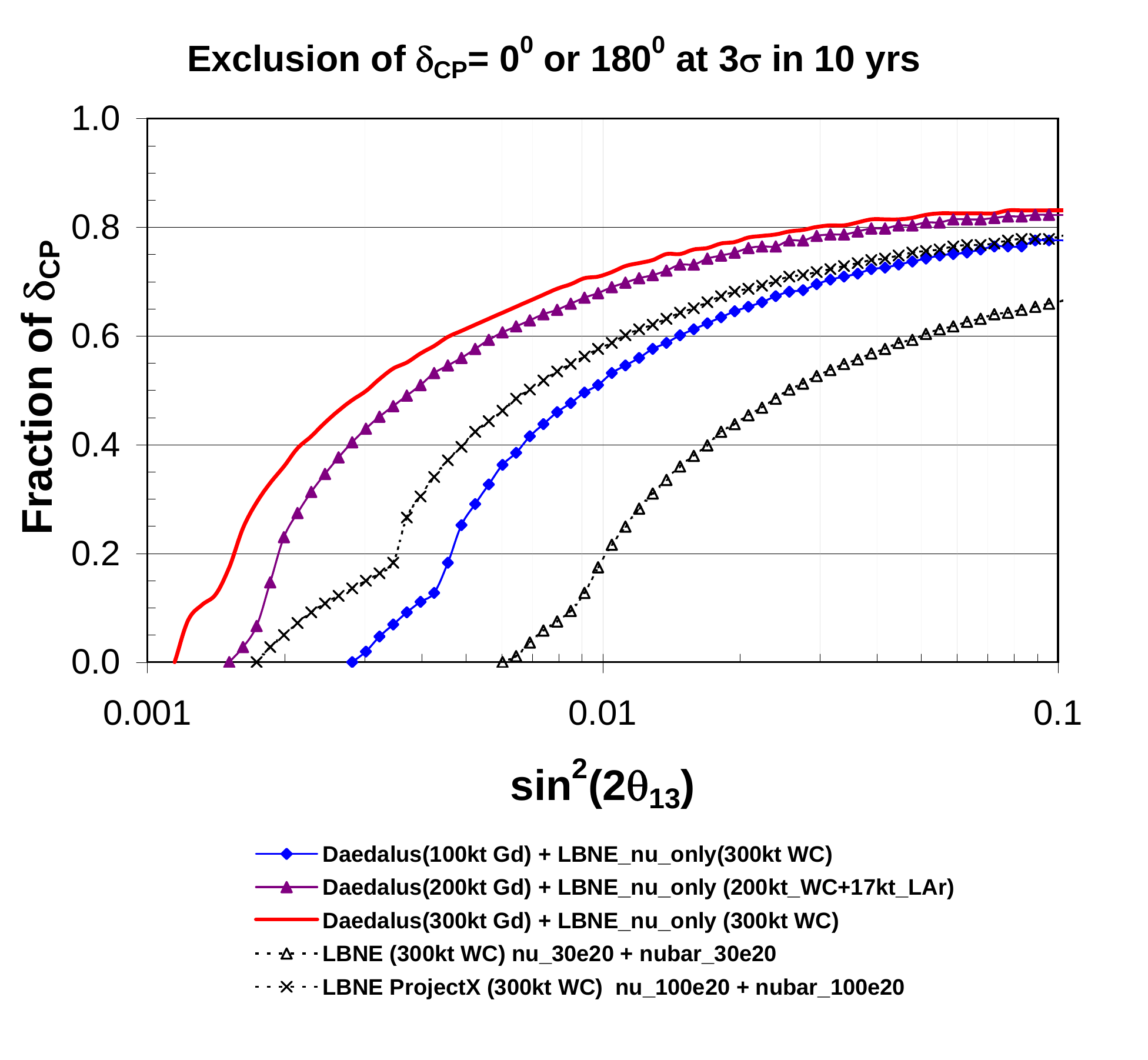}} 
\end{center}
\caption{The fraction of $\delta_{CP}$-space which excludes $0^\circ$
  or $180^\circ$ at 3$\sigma$ for various configurations involving
  units of LAr, WC and WCGd assuming DAE$\delta$ALUS plus LBNE
  $\nu-$only scenarios for 10 years of running.  Three examples,
  ranked 6, 9 and 10 in Table 1, with 100 kt (blue diamond), 200 kt
  (purple triangle) and 300 kt of WCGd (bold red line), are presented.
  In each case, the LBNE running is with three modules, either 3 WC or
  2 WC plus 1 LAr -- thus LBNE events come from 300 kt-equivalent in
  all three cases.  These are compared with the Project X sensitivity
  (black x) and the standard LBNE-only run, both of which employ 5
  years of $\nu$ plus 5 years $\bar{\nu}$ running with three modules,
  300 kt WC.  }
\end{figure}

\newpage

\vspace{0.5in}
{\footnotesize

\hrule

\vspace{0.1in}

\noindent $^*$DAE$\delta$ALUS Authors:

\vspace{0.1in}

\noindent J. Alonso$^{13}$, F.T. Avignone$^{18}$, W.A. Barletta$^{13}$,
R. Barlow$^5$, H.T. Baumgartner$^{13}$, A. Bernstein$^{11}$, E. Blucher$^4$,
L. Bugel$^{13}$, L. Calabretta$^9$, L. Camilleri$^6$, R. Carr$^6$, 
J.M. Conrad$^{13,*}$, S.A. Dazeley$^{11}$, Z. Djurcic$^2$, A.~de~Gouv\^ea$^{17}$,
P.H. Fisher$^{13}$, C.M. Ignarra$^{13}$, B.J.P. Jones$^{13}$, C.L. Jones$^{13}$,G. Karagiorgi$^{13}$, T. Katori$^{13}$, S.E. Kopp$^{20}$, R.C. Lanza$^{13}$,
W.A. Loinaz$^1$, P. McIntyre$^{19}$, G. McLaughlin$^{16}$, G.B. Mills$^{12}$,
J.A. Nolen$^2$, V. Papavassiliou$^{15}$, M. Sanchez$^{2,10}$,K. Scholberg$^7$,
W.G. Seligman$^6$, M.H. Shaevitz$^{6,*}$, S. Shalgar$^{17}$, T. Smidt$^{13}$,
M.J. Syphers$^{14}$, J. Spitz$^{22}$, H.-K. Tanaka$^{13}$, K. Terao$^{13}$,
C. Tschalaer$^{13}$, M. Vagins$^{3,21}$, R. Van de Water$^{12}$,  
M.O. Wascko$^8$, R. Wendell$^7$, L. Winslow$^{13}$

\vspace{0.2in}

\noindent $^1$Amherst College, Amherst, MA 01002, USA \\
$^2$Argonne National Laboratory, Argonne, IL 60439, USA \\
$^3$University of California, Irvine, CA 92697, USA \\ 
$^4$University of Chicago, Chicago, IL 60637, USA \\
$^5$The Cockcroft Institute for Accelerator Science \& \\
the University of Manchester, Oxford Road, Manchester M13 9PL, UK\\
$^6$Columbia University, New York, NY 10027, USA\\
$^7$Duke University, Durham, NC 27708, USA \\
$^8$Imperial College London. London, SW7 2AZ, UK\\
$^9$Istituto Nazionale di Fisica Nucleare, Laboratori Nazionali del Sud, I-95123, Italy\\
$^{10}$Iowa State University, Ames, IA 50011, USA\\
$^{11}$Lawrence Livermore National Laboratory, Livermore, CA 94551, USA \\
$^{12}$Los Alamos National Laboratory, Los Alamos, NM 87545, USA \\
$^{13}$Massachusetts Institute of Technology, Cambridge, MA 02139, USA\\
$^{14}$Michigan State University, East Lansing, MI 48824, USA\\
$^{15}$New Mexico State University, Las Cruces, NM 88003, USA\\
$^{16}$North Carolina State University, Raleigh, NC 27695, USA\\
$^{17}$Northwestern University, Evanston, IL 60208, USA \\
$^{18}$University of South Carolina, Columbia, SC 29208,USA \\
$^{19}$Texas A\&M University, College Station, TX 77843, USA \\
$^{20}$University of Texas, Austin, TX 78712, USA\\
$^{21}$University of Tokyo, Kashiwa, 277-8583, Japan\\
$^{22}$Yale University, New Haven, CT 06520 USA \\}

\end{document}